\documentclass[aps,twocolumn,superscriptaddress,showkeys,showpacs]{revtex4}
\usepackage{amssymb}
\usepackage{amsfonts}
\usepackage{amsmath}
\usepackage{amsthm}
\usepackage{epsfig}
\usepackage{color} 
\usepackage{subfigure}
\usepackage{bbm}
\usepackage{comment}

\begin{document}
\title{Deterministic and stochastic coarsening control in optically-addressed spatial light modulators subject to optical feedback}

\author{Vladimir V. Semenov}
\email[corresponding author: ]{semenov.v.v.ssu@gmail.com}
\affiliation{Institute of Physics, Saratov State University, Astrakhanskaya str. 83, 410012 Saratov, Russia}
\affiliation{FEMTO-ST Institute/Optics Department, UMR CNRS 6174, University Franche-Comt\'e, \\15B avenue des Montboucons,
Besan\c con Cedex, 25030, France}

\author{Xavier Porte}
\affiliation{FEMTO-ST Institute/Optics Department, UMR CNRS 6174, University Franche-Comt\'e, \\15B avenue des Montboucons,
Besan\c con Cedex, 25030, France}
\affiliation{Institute of Photonics, Department of Physics, University of Strathclyde, 99 George str., Glasgow G1 1RD, UK}

\author{Laurent Larger}
\affiliation{FEMTO-ST Institute/Optics Department, UMR CNRS 6174, University Franche-Comt\'e, \\15B avenue des Montboucons,
Besan\c con Cedex, 25030, France}

\author{Daniel Brunner}
\affiliation{FEMTO-ST Institute/Optics Department, UMR CNRS 6174, University Franche-Comt\'e, \\15B avenue des Montboucons,
Besan\c con Cedex, 25030, France}

\date{\today}

\begin{abstract}
Phase separation accompanied by further domain growth and coarsening is a phenomenon common to a broad variety of dynamical systems. In this connection, controlling such processes represents a relevant interdisciplinary problem. Using methods of numerical modelling, we demonstrate two approaches for the coarsening control in bistable systems based on the example of a spatially-extended model describing an optically-addressed spatial light modulator with two color illumination subject to optical feedback. The first method implies varying system parameters such that the system evolves as the pitchfork or saddle-node normal forms. The second method leverages noise whose intensity is used as an additional system parameter. Both, deterministic and stochastic schemes allow to control the direction and speed of the fronts separating spatial domains. The considered stochastic control represents a particular case of the noise-sustained front propagation in bistable systems and involves the properties of the optical system under study. In contrast, the proposed deterministic control technique can be applied to bistable systems of different nature.
\end{abstract}
\pacs{05.10.-a, 05.40.-a, 43.50.+y, 46.65.+g}
\keywords{spatially-extended system, bistability, front propagation, coarsening, noise, control, pitchfork bifurcation, saddle-node bifurcation, bifurcation normal forms}
\maketitle

\section{Introduction}
Besides the well-known Turing patterns, reaction-diffusion systems exhibit a big variety of spatio-temporal dynamics \cite{kuramoto1984,mikhailov1990,kapral1995,garcia-ojalvo1999} including traveling fronts, solitary and periodic pulses, spiral turbulence, scroll waves and noise-induced pattern formation. In particular, bistable reaction-diffusion media can exhibit dynamics, where for the case when two kinds of domains form and evolve in space, separating fronts between them arise and propagate. Such propagating wavefronts \footnote{It is important to note that the term 'wavefront' does not refer to the surface over which an optical wave has a constant phase. Here, a wavefront or simply a front refers to the context of, for example, fluid dynamics and describes a boundary between domains corresponding to different quiescent steady state regimes in bistable reaction-diffusion systems.} are of frequent occurrence in chemistry, see, for instance, the Schl{\"o}gl model \cite{schloegl1972,schloegl1983,loeber2014} developed for the explanation of an autocatalytic reaction mechanism, as well as in electronics \cite{schoell2001}, flame propagation theory \cite{zeldovich1938}, just to name a few. 

In the simplest case, front propagation appears in 1D-space. If a studied bistable media evolves in 2D-space, then the peculiarities of front propagation are additionally determined by the shape of domains formed by such fronts. In such a case, one observes an effect often referred to as 'coarsening'. Coarsening is a particular form of front propagation, and corresponds to the expansion of domains that invade the entire space on the cost of other domains. It is a fundamental phenomenon demonstrated in the context of different areas: physics of liquid crystals \cite{yurke1992} and magnetism \cite{bray1994,cugliandolo2010,caccioli2008,denholm2019}, physics and chemistry of materials \cite{goh2002,zhang2019,zhang2019-2,geslin2019}, laser physics \cite{yanchuk2012,marino2014,javaloyes2015}, electronics \cite{semenov2018} and animal population statistics \cite{dobramysl2018}. It occurs in bistable spatially-extended systems \cite{bray1994} and time-delay oscillators \cite{yanchuk2012,marino2014,semenov2018} when a bistable system is prepared in an inhomogeneous state including both steady states. Under such conditions, separating fronts start to propagate such that growing spatial domains appear and one state (phase) therefore starts to dominate the whole system. Analogous processes can arise in stochastic systems as an accompaniment of noise-induced phase transitions \cite{vandenbroeck1997,carrillo2003}.
 
 %%%%%%%%%%%%%%%%%%%%%%% FIG 1 %%%%%%%%%%%%%%%%%%%%%%%%
\begin{figure*}[t!]
\centering
\includegraphics[width=0.65\textwidth]{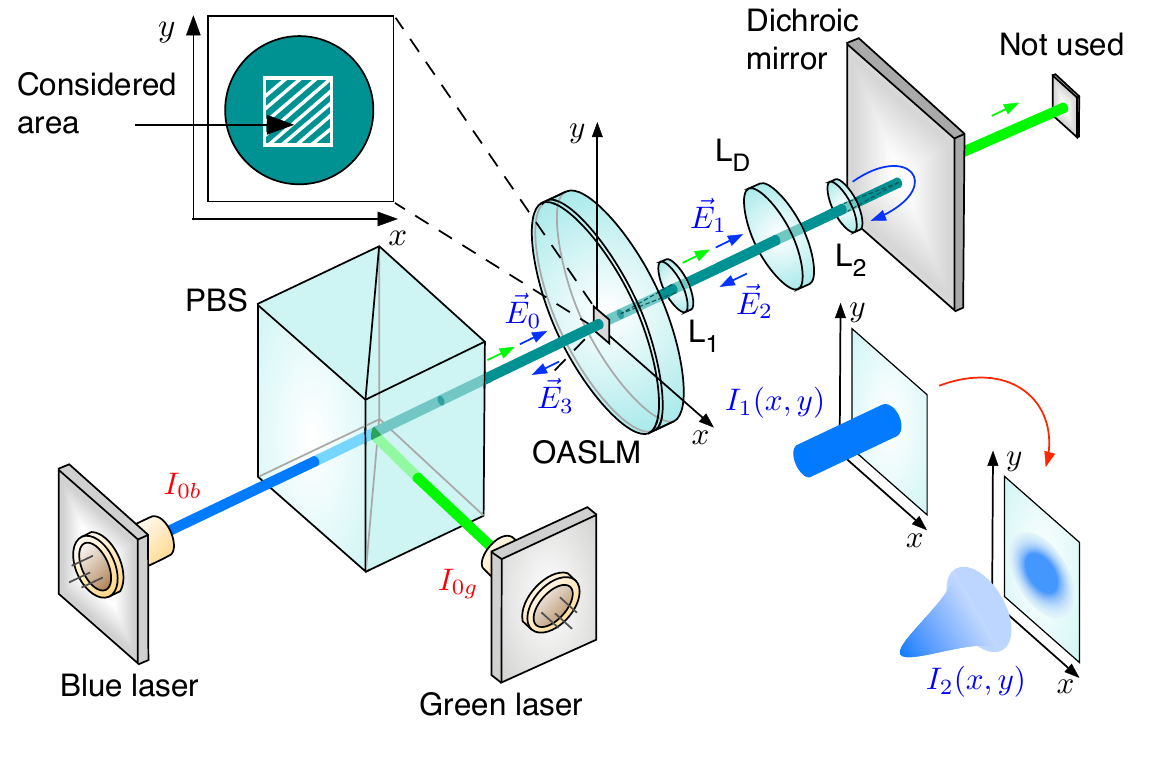}\\
\caption{Single-PS-layer OASLM under simultaneous blue and green illumination when the blue light beam is reflected by the dichroic mirror and creates feedback. The system contains a defocusing lens to emulate local diffusion by spatially broadening the field distribution of the back-reflected optical field. Lenses L1 and L2 create 4f-imaging of the OASLM's state back on itself after reflection by the mirror.}
\label{fig1}
\end{figure*}  
%%%%%%%%%%%%%%%%%%%%%%%%%%%%%%%%%%%%%%%%%%%%%%%%%%%
 
It is well-known that the presence of any kind of asymmerty in bistable spatially-extended systems has a principal impact on the speed of wavefront propagation, for instance, in bistable reaction-diffusion models \cite{engel1985,loeber2014}:  the bigger is the asymmetry, the faster wavefronts propagate. Moreover, control over the system's asymmetry allows to stop the wavefront propagation or to even invert its direction. In the current paper we illustrate these facts by means of numerical modelling on an  example of a spatially-extended bistable dynamical system describing the optical device considered in Ref. \cite{semenov2021}. In particular, we use the Taylor-series-based technique of the pitchfork and saddle-node bifurcation implementation developed in Ref. \cite{semenov2021} to control the system's asymmetry and consequently to control coarsening. 

The second strategy for controlling the propagating fronts is based on noise \cite{schimansky1983,engel1985,loecher1998}. In particular, it has been established that multiplicative noise can influence the systematic part of front dynamics \cite{engel1985,garcia-ojalvo1999,mendez2011}. We numerically show how the stochastic scheme of the propagation control can be implemented in terms of optics on the example of the optical device stochastic model.

Generally speaking, we consider two options for coarsening control in bistable dynamical systems: deterministic and stochastic approaches. The deterministic control scheme is a universal methodology for approaching the model equations to the pitchfork or saddle-node bifurcation normal forms and can be applied to dynamical systems of different nature. In contrast, the stochastic approach involves the specificity of the concrete optical device model and hence is not universal. However, the obtained results complement a manifold of stochastic phenomena associated with propagating fronts by optical processes and are potentially interesting for specialists in optics, nonlinear dynamics and theory of stochastic processes. 

\section{Model under study}
The central element of the system discussed in this paper (see Fig. \ref{fig1}) is an optically-addressed spatial light modulator (OASLM). Our OASLM model was developed in \cite{semenov2021} based on an experimental study of the OASLM sample fabricated according to the concept reported in \cite{kirzhner2014}. The OASLM is a light-transmissive device, and it is assumed in the following that the OASLM fully transmits the incident light, i.e. has zero absorption. This is a valid approximation, as the devices relies on nanometer thin photo-sensitive layers. The OASLM operates as an optically controlled birefringent phase plate leveraging a nematic liquid crystal (LC) layer, the phase retardation of which, $\Gamma$, is a dynamic quantity. The LC-layer is located between two two a-As$_2$S$_3$ chalcogenide thin films that simultaneously function as a photosensitive (PS) and alignment layers. The OASLM is connected to a DC-power source resulting in a voltage drop across the LC layer that is uniform without illumination. However, illumination spatially modifies the PS’s conductivity and in turn the local voltage drop across the LC layer. Due to the induced dipole moment, LC molecules change their orientation in response, which results in a spatial birefringence distribution that is a function of the optical illumination profile. Consequently, an optical wave crossing the LC experiences a change of its polarization state due to phase retardation $\Gamma$ between the ordinary and the extraordinary axes and $\Gamma(I)=(\alpha I +\beta)^{-1}+\gamma$ \cite{semenov2021}. Noteworthy, the second PS-layer does not increase the device's dynamical complexity and has no principal impact on the dynamics, besides doubling the responsivity of the device \cite{semenov2021}. In the rest of the manuscript we therefore only consider an OASLM with a single PS-layer located on the left side of the LC-layer. 

In our generic setup, depicted in Fig. \ref{fig1}, the single-PS-layer OASLM operates in the phase modulation regime (the OASLM's rotation angle $\psi=m\pi$ where $m\in \mathbb{Z}$). After traversing the OASLM, a dichroic mirror transmits the green light beam but reflects the blue with reflectivity $R$, which then interferes with the original blue illumination. Thus, the dichroic mirror creates optical feedback and potentially coupling. Generally, optical interference, i.e. a temporal beating originating from the superposition blue and green light, can be ignored due to the vast difference in frequencies of both light sources. It is assumed that the PS's thickness is significantly smaller than the wavelength. 

Using Jones matrix calculus, we describe the optical field distributions which determine the system dynamics. Here, we consider uniformly distributed horizontally polarized blue illumination, $\vec{E}_0(x,y)=[E_0,0]$. After passing the OASLM one obtaines $\vec{E}_1(x,y)~=~\exp\left(i(\phi_0+\Gamma(x,y))\right)\vec{E}_0(x,y)$, where $\phi_0$ is the constant retardation induced by the OASLM without illumination and $i$ is the imaginary unit. The setup in Fig. \ref{fig1} contains two optical lenses L$_1$ and L$_2$ to create 4f-imaging of the OASLM's state back on itself after reflection by the mirror, and a defocusing lens L$_D$ within the optical feedback path. Defocusing leads to blurred imaging, as illustrated in Fig. \ref{fig1}, and its impact can be mathematically described as a convolution with a Gaussian of a width that can be controlled through the positioning lenses. Applying this, one obtains a spatial distribution of the returned light Jones vector $\vec{E}_{2}(x,y)$ as 
%\begin{equation}
%\begin{array}{l}
\begin{widetext}
\begin{equation}
\label{E_2_convolutional}
\vec{E}_{2}(x,y)=\Bigg(R\exp(\phi_1)\vec{E}_1(x,y)\Bigg)
\ast \Bigg( \dfrac{1}{2\pi\sigma^2}\exp\left(-\dfrac{x^2}{2\sigma^2}-\dfrac{y^2}{2\sigma^2} \right)\Bigg),
\end{equation}
\end{widetext}
%\end{array}
%\end{equation}
where $\phi_1$ is the retardation accumulated in the external cavity round-trip, '$\ast$' is the convolution operation and the Gaussian function plays a role of a point spread function widened from the normal imaging setup via the defocusing lens. Finally, the optical wave passes through the OASLM again and one obtains field $\vec{E}_3(x,y)~=~\exp\left(i(\phi_0+\Gamma(x,y))\right)\vec{E}_2(x,y)$ the resulting blue light field at the PS-layer on the left side of the OASLM is $\vec{E}_{\text{b}}(x,y)=\vec{E}_{0}(x,y)+\vec{E}_{3}(x,y)$ with intensity $I_{\text{b}}(x,y)=\left| \vec{E}_{\text{b}}(x,y)\right|^2$. To simplify the model, diffusive processes inside the OASLM are neglected and width $\sigma$ of the optical convolution kernel is chosen to be several times greater than the OASLM resolution, $\sigma_{\text{OASLM}}=3.5 \mu$m (see details in papers \cite{kirzhner2014,semenov2021}). Finally, optical feedback is considered instantaneous relative to the OASLM’s response time $\varepsilon$. The temporal evolution of the blue light's retardation then takes the form
\begin{widetext}
\begin{equation}
\label{spatial_model_phase_modulation_conv}
\begin{array}{l}
\vec{E}_0(x,y)~=~
\begin{bmatrix}
E_{0}\\
0
\end{bmatrix}, \quad\quad

\vec{E}_1(x,y)~=~\exp\left(i(\phi_0+\Gamma(x,y))\right)
\begin{bmatrix}
E_0 \\
0
\end{bmatrix}
,\\
\vec{E}_{2}(x,y)=\Bigg(R\exp(\phi_1)\vec{E}_1(x,y)\Bigg)\ast \Bigg( \dfrac{1}{2\pi\sigma^2}\exp\left(-\dfrac{x^2}{2\sigma^2}-\dfrac{y^2}{2\sigma^2} \right)\Bigg),\\
\vec{E}_3(x,y)~=~\exp\left(i(\phi_0+\Gamma(x,y))\right)\vec{E}_2(x,y),\quad\quad I_{\text{b}}(x,y)= \left| \vec{E}_{0}(x,y)+\vec{E}_{3}(x,y) \right|^2,\\
\varepsilon \dfrac{d\Gamma(x,y)}{dt}=-\Gamma(x,y)+\dfrac{1}{\alpha_{\text{b}} I_{\text{b}}(x,y)+ \tilde{\alpha}_{\text{g}} I_{\text{0g}}+\beta}+\gamma,
\end{array}
\end{equation}
\end{widetext}
where $\tilde{\alpha}_{\text{g}} = \frac{\lambda_{\text{g}}}{\lambda_{\text{b}}}\alpha_g$ is the retardation effect of $I_{0\text{g}}$ on the blue signal. 

If one neglects spatial aspects of the dynamics and excludes the optical convolution implemented by L$_D$, the system is reduced into an ordinary differential equation (see paper \cite{semenov2021}) for the temporal evolution of the retardation:
\begin{equation}
\label{single_oscillator}
\begin{array}{l}
\varepsilon \dfrac{d\Gamma}{dt}=-\Gamma+\dfrac{1}{\alpha_{\text{b}} I_{\text{b}}(\Gamma)+\tilde{\alpha}_g I_{0\text{g}}+\beta}+\gamma,\\
I_{\text{b}}(\Gamma)=I_{0\text{b}}\Big\{1+R^2+ 2 R\cos(2\phi_0+\phi_1+2\Gamma)       \Big\}.
\end{array}
\end{equation}
 
The action of the convolution operation in spatially-extended model (\ref{spatial_model_phase_modulation_conv}) is associated with homogenous coupling of the system state at any point on the plane ($x$,$y$) with its neighbour states in some range $x\in[x-\Delta x; x+\Delta x]$, $y\in[y-\Delta y; y+\Delta y]$. Defocusing represents a natural physical approach for the homogeneous coupling  implementation similarly to diffusive effects occurring inside the OASLM. If the coupling radius does not exceed the OASLM's linear pixel size, one deals with local coupling, whose impact is identical to the action of diffusion. Then one can expect to observe the effects of wave propagation and coarsening in Eq. (\ref{spatial_model_phase_modulation_conv}), where the system parameters correspond to the regime of bistability in single-oscillator of Eq. (\ref{single_oscillator}). Our numerical study is based on modelling Eq. (\ref{spatial_model_phase_modulation_conv}) using the Heun method \cite{mannella2002} with time step $\Delta t=10^{-3}$. In the rest of the manuscript Eq. (\ref{spatial_model_phase_modulation_conv}) are studied for the fixed parameter set 
$R=0.95$, $\alpha_{\text{b}}=0.117$ [m$^2$Rad$^{-1}$W$^{-1}$], $\alpha_{\text{g}}=98.5\times 10^{-6}$ [m$^2$Rad$^{-1}$W$^{-1}$], $\beta=0.052$ [Rad$^{-1}$], $\gamma=-0.55$ [Rad], $\phi_0=\pi/2$ [Rad], $\phi_1=\pi$ [Rad], $\varepsilon=1$ [s], $\sigma=10^{-5}$ [m], and varying $I_{0\text{b}}$ and $I_{0\text{g}}$. The blue and green light wavelength are chosen according to experiments carried out in article \cite{semenov2021}, $\lambda_{\text{b}}=450\times10^{-9}$ [m] and $\lambda_{\text{g}}=532\times10^{-9}$ [m] correspondingly.

\section{Deterministic control}
\begin{figure}[t!]
\centering
\includegraphics[width=0.5\textwidth]{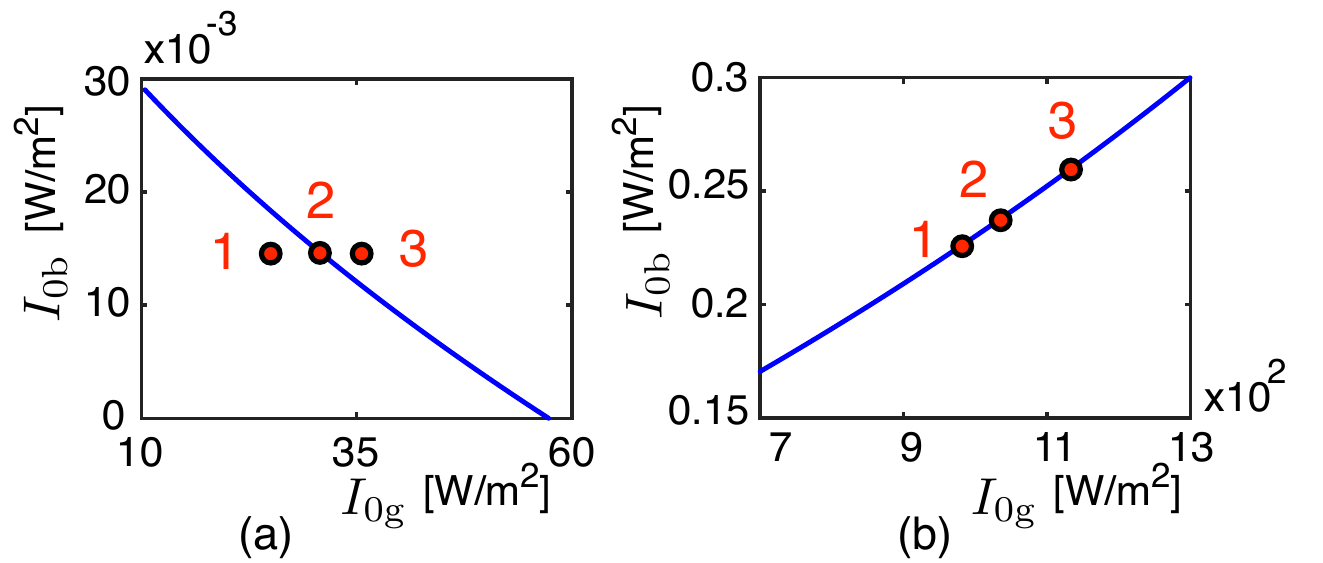}\\
\caption{Pitchfork (panel (a)) and saddle-node (panel (b)) bifurcation condition curves for varying $I_{0\text{b}}$ and $I_{0\text{g}}$ in Eq. (\ref{single_oscillator}) (see Ref. \cite{semenov2021} for details). System parameters are:  $\varepsilon=1$ [s], $\alpha_{\text{b}}=0.117$ [m$^2$Rad$^{-1}$W$^{-1}$], $\alpha_{\text{g}}=98.5\times 10^{-6}$ [m$^2$Rad$^{-1}$W$^{-1}$], $\beta=0.052$ [Rad$^{-1}$], $\gamma=-0.55$ [Rad], $\phi_0=\pi/2$ [Rad], $\phi_1=\pi$ [Rad], $\lambda_{\text{b}}=450\times10^{-9}$ [m], $\lambda_{\text{g}}=532\times10^{-9}$ [m], $R=0.95$.}
\label{fig2}
\end{figure}

As demonstrated in \cite{semenov2021}, one can implement the pitchfork and the saddle-node bifurcation of steady states in our system described by Eq. (\ref{single_oscillator}), if $I_{0\text{b}}$ and $I_{0\text{g}}$ are chosen according to the corresponding bifurcation condition curves depicted in Fig. \ref{fig2}. In more detail, when $I_{0\text{b}}$ and $I_{0\text{g}}$ are varied according to the curves in Fig. \ref{fig2}, the right-hand side function $f(\Gamma)$ of Eq. (\ref{single_oscillator}) represented in the form $\dfrac{d\Gamma}{dt}=f(\Gamma)$ evolves in some range of $\Gamma$ as a cubic and a quadratic function. Then Eq. (\ref{single_oscillator}) is considered as the pitchfork and saddle-node bifurcation normal forms, $\dfrac{d\Gamma}{dt}=b\Gamma-d\Gamma^3$ and $\dfrac{d\Gamma}{dt}=a+c\Gamma^2$. Here, we use these bifurcation conditions to control the effect of coarsening in Eq. (\ref{spatial_model_phase_modulation_conv}). We consider Eq. (\ref{single_oscillator}) in the form $\dfrac{d\Gamma}{dt}=f(\Gamma)$ and illustrate the right-hand side function $f(\Gamma)$ for varying $I_{0\text{b}}$ and $I_{0\text{g}}$. For all parameter values, $f(\Gamma)$ shows three steady states corresponding to the condition $f(\Gamma)=0$: stable steady states A and B and an unstable equilibrium between them (see Fig. \ref{fig3} and Fig. \ref{fig4}). We visualise the fact that the symmetry properties of Eq. (\ref{single_oscillator}) describing the local dynamics without coupling are reflected in the duration and direction of coarsening in Eq. (\ref{spatial_model_phase_modulation_conv}).  For this purpose, we fix the initial spatial pattern at $t=0$ and observe the spatial evolution when $I_{0\text{b}}$ and $I_{0\text{g}}$ change.
%%%%%%%%%%%%%%%%%%%%%%% FIG 2 %%%%%%%%%%%%%%%%%%%%%%%%
%%%%%%%%%%%%%%%%%%%%%%%%%%%%%%%%%%%%%%%%%%%%%%%%%%%
\subsection{Pitchfork bifurcation conditions}
%%%%%%%%%%%%%%%%%%%%%%% FIG 3 %%%%%%%%%%%%%%%%%%%%%%%%
\begin{figure*}[t]
\centering
\includegraphics[width=0.9\textwidth]{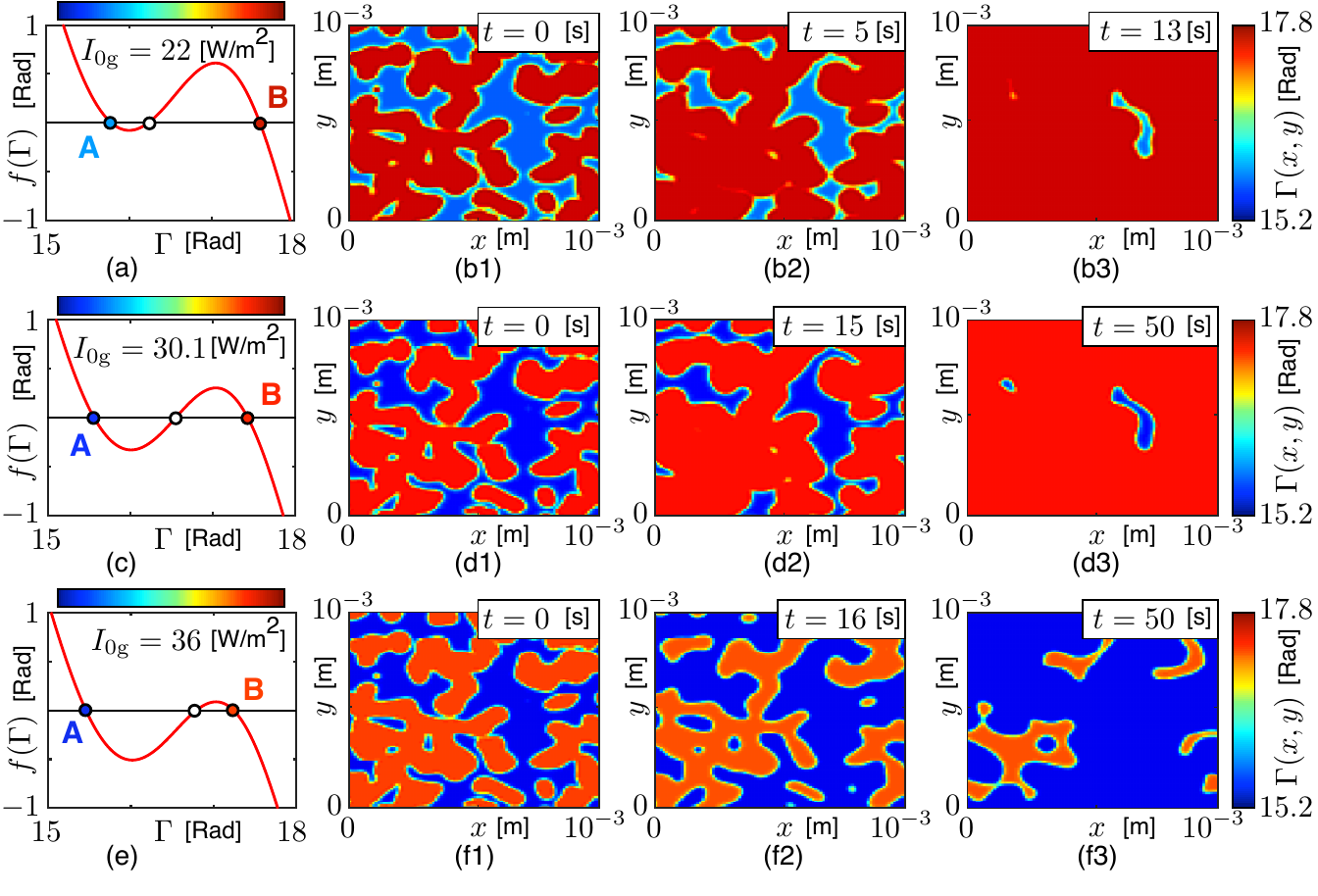}\\
\caption{Coarsening and the pitchfork bifurcation conditions: evolution of the right-hand side function of Eq. (\ref{single_oscillator}) and coarsening in Eq. (\ref{spatial_model_phase_modulation_conv}) for increasing green light intensity: $I_{0\text{g}}=22$  [W/m$^2$] (panels (a) and (b)) corresponding to point 1 in Fig. \ref{fig2} (a), $I_{0\text{g}}=30.1$ [W/m$^2$] (panels (c) and (d)) corresponding to point 2 in Fig. \ref{fig2} (a), $I_{0\text{g}}=36$ [W/m$^2$] (panels (e) and (f)) corresponding to point 3 in Fig. \ref{fig2} (a). Other parameters are:  $\varepsilon=1$ [s], $\alpha_{\text{b}}=0.117$ [m$^2$Rad$^{-1}$W$^{-1}$], $\alpha_{\text{g}}=98.5\times 10^{-6}$ [m$^2$Rad$^{-1}$W$^{-1}$], $\beta=0.052$ [Rad$^{-1}$], $\gamma=-0.55$ [Rad], $\phi_0=\pi/2$ [Rad], $\phi_1=\pi$ [Rad], $\lambda_{\text{b}}=450\times10^{-9}$ [m], $\lambda_{\text{g}}=532\times10^{-9}$ [m], $R=0.95$, $I_{0\text{b}}=0.01506$ [W/m$^2$], $\sigma=10^{-5}$ [m].}
\label{fig3}
\end{figure*}
%%%%%%%%%%%%%%%%%%%%%%%%%%%%%%%%%%%%%%%%%%%%%%%%%%%

Let us fix light intensities $I_{0\text{g}}=22$, $I_{0\text{b}}=|\vec{E}_0|^2=0.015$ [W/m$^2$] in Eq. (\ref{spatial_model_phase_modulation_conv}). This parameter set corresponds to the regime of bistability in the single-oscillator model described by Eq. (\ref{single_oscillator}), but the pitchfork bifurcation conditions are not fulfilled (this parameter set corresponds to point 1 in Fig.~\ref{fig2}~(a)) and the right-hand side function of Eq. (\ref{single_oscillator}) is asymmetric, see Fig. \ref{fig3} (a). In that case, the spatially extended model described by Eq. (\ref{spatial_model_phase_modulation_conv}) exhibits coarsening, see Fig. \ref{fig3} (b1-b3). The system asymmetry is reflected in the fact that the basin of attraction of state B is larger than the one of state A, and the unstable fixed point is closer to attractor A than to the stable steady state B. This results in the spatial evolution of Eq.  (\ref{spatial_model_phase_modulation_conv}) such that the red domains corresponding to state B extend and invade the entire space ($x$,$y$). 

Increasing $I_{0\text{g}}$ allows to fulfil the pitchfork bifurcation conditions at $I_{0\text{g}}\approx30.1$ [W/m$^2$] (point 2 in Fig.~\ref{fig2}~(a)), for which the asymmetry of the right-hand side function $f(\Gamma)$ is minimized, see Fig. \ref{fig3} (c), and coarsening is substantially slower. Consequently, a longer time is necessary for the transformation of the same initial metastable state as in Fig. \ref{fig3} (b1) (the initial states in Fig. \ref{fig3}(b1,d1,f1) are identical) into the quiescent regime when either steady state A or B invades the entire space, see Fig. \ref{fig3} (d1-d3). It should be noted that in the case of minimal asymmetry, the probabilities to observe the final state $\Gamma(x,y)=A$ or $\Gamma(x,y)=B$ starting from random initial conditions are almost identical.  

If one continues to increase the green light intensity, the phase space structure is inverted in comparison with the initial configuration, as can be seen from comparison of $f(\Gamma)$ in Fig. \ref{fig3} (a,e). The motion of fronts separating domains reverses, and coarsening direction becomes opposite: steady state A invades the whole space, see Fig. \ref{fig3}(f1-f3) corresponding to $I_{0\text{b}}=0.015$ [W/m$^2$] and $I_{0\text{g}}=36$ [W/m$^2$] (point 3 in Fig. \ref{fig2} (a)).

\subsection{Saddle-node bifurcation conditions}
\begin{figure*}[t]
\centering
\includegraphics[width=0.9\textwidth]{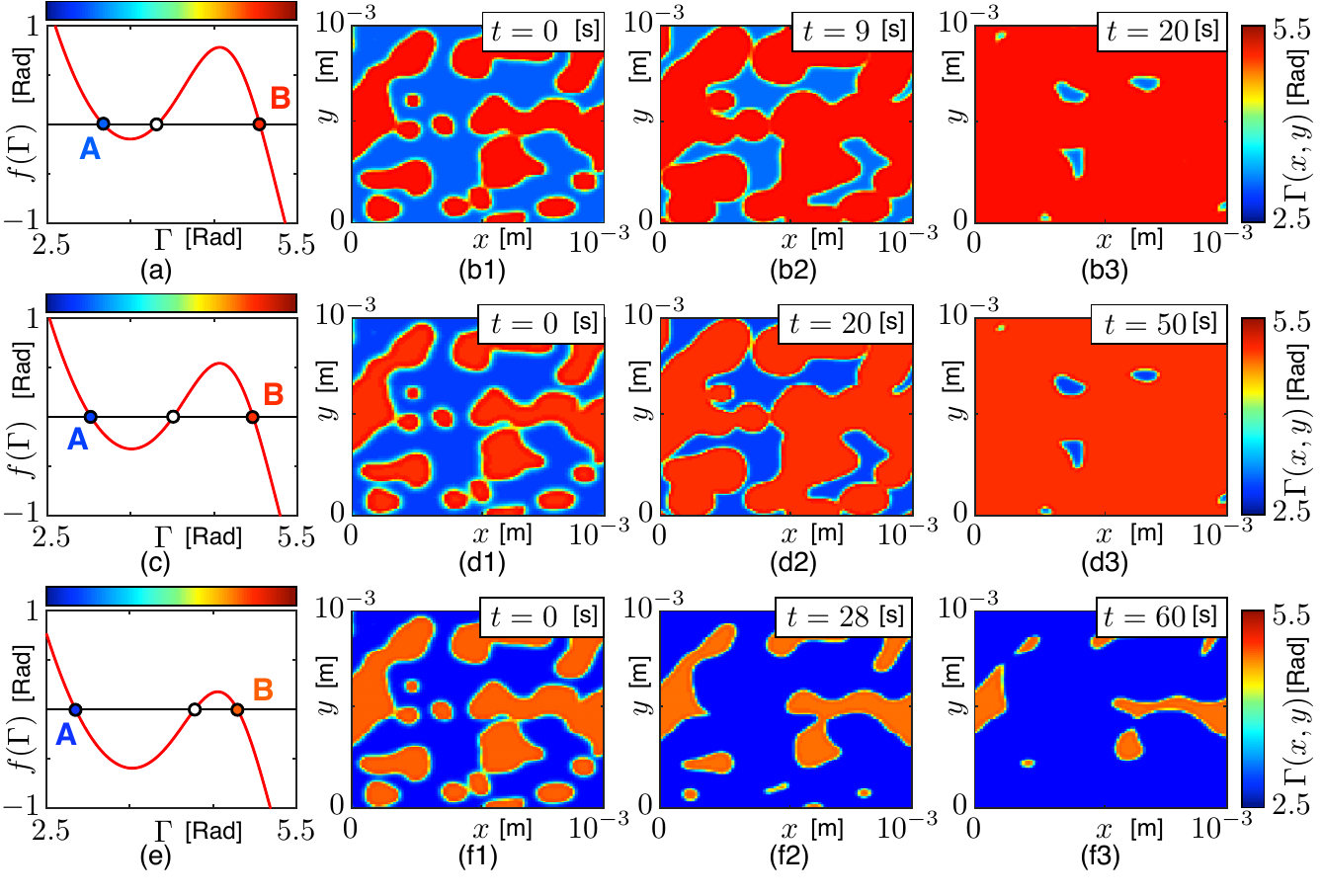}\\
\caption{Coarsening and the saddle-node bifurcation conditions: evolution of the right-hand side function of Eq. (\ref{single_oscillator}) and coarsening in Eq. (\ref{spatial_model_phase_modulation_conv}) when $I_{0\text{b}}$ and $I_{0\text{g}}$ vary according to the saddle-node bifurcation conditions for Eq. (\ref{single_oscillator}): $I_{0\text{b}}=0.228$ [W/m$^2$], $I_{0\text{g}}=990$ [W/m$^2$] (panels (a) and (b)) corresponding to point 1 in Fig. \ref{fig2} (b), $I_{0\text{b}}=0.241$ [W/m$^2$], $I_{0\text{g}}=1050$ [W/m$^2$] (panels (c) and (d)) corresponding to point 2 in Fig. \ref{fig2} (b), $I_{0\text{b}}=0.2645$ [W/m$^2$], $I_{0\text{g}}=1153$ [W/m$^2$] (panels (e) and (f)) corresponding to point 3 in Fig. \ref{fig2} (b). Other parameters are:  $\varepsilon=1$ [s], $\alpha_{\text{b}}=0.117$ [m$^2$Rad$^{-1}$W$^{-1}$], $\alpha_{\text{g}}=98.5\times 10^{-6}$ [m$^2$Rad$^{-1}$W$^{-1}$], $\beta=0.052$ [Rad$^{-1}$], $\gamma=-0.55$ [Rad], $\phi_0=\pi/2$ [Rad], $\phi_1=\pi$ [Rad], $\lambda_{\text{b}}=450\times10^{-9}$ [m], $\lambda_{\text{g}}=532\times10^{-9}$ [m], $R=0.95$, $\sigma=10^{-5}$ [m].}
\label{fig4}
\end{figure*}

Varying $I_{0\text{b}}$ and $I_{0\text{g}}$ according to the curve obtained using the saddle-node bifurcation conditions, corresponding to the blue line in Fig.~\ref{fig2}~(b), allows to move the right-hand side function of Eq. (\ref{single_oscillator}) up and down, see Fig.~\ref{fig4}~(a,c,e). A symmetric configuration of $f(\Gamma)$ can be achieved during this motion, see Fig.~\ref{fig4}~(c), and the same effects as in the previous section can be observed. First, the system's asymmetry is well-pronounced, as illustrated in Fig.~\ref{fig4}~(a), and the state B rapidly invades the space ($x$,$y$), see Fig.~\ref{fig4}~(b1-b3). When $I_{0\text{b}}$ and $I_{0\text{g}}$ are adjusted such that the saddle-node bifurcation conditions are fulfilled, the system passes through the symmetric state (see Fig.~\ref{fig4}~(c)), and the coarsening effect slows down as illustrated in Fig.~\ref{fig4}~(d1-d3). Further changing $I_{0\text{b}}$ and $I_{0\text{g}}$ inverts the asymmetry, see Fig.~\ref{fig4}~(e), and the motion of fronts separating blue and red domains reverses its direction, see Fig.~\ref{fig4}~(f1-f3). 

%%%%%%%%%%%%%%%%%%%%%%% FIG 4 %%%%%%%%%%%%%%%%%%%%%%%%
%%%%%%%%%%%%%%%%%%%%%%%%%%%%%%%%%%%%%%%%%%%%%%%%%%%

\section{Stochastic control}
Consider a stochastic model of the optical setup illustrated in Fig. \ref{fig1}. For that purpose, it is assumed that the green light illumination contains a stochastic contribution according to $I_{0\text{g}}(x,y)=I_{0\text{g}}+\xi(x,y)$. Here, $\xi(x,y)$ represents a source of spatial coloured noise described by the first-order Ornstein-Uhlenbeck process 
 \begin{equation}
 \label{ornstein-uhlenbeck}
 \tau_{c}\dfrac{d\xi(x,y)}{dt}=-\xi(x,y)+\sqrt{2D_{\text{g}}\tau_{c}} n(x,y,t), 
 \end{equation}
 where $\tau_{c}$ is the coloured noise correlation time, $n(x,y,t)$ is a normalized source of white Gaussian noise, $D_{\text{g}}$ plays a role of the noise intensity. The temporal and spatial correlation properties of the noise source $n(x,y,t)$ at any point $\vec{r}_0$ are described by the delta function $<n(\vec{r}_0,t)>=0$, $<n(\vec{r}_0,t)n(\vec{r}_0,t+\tau)>= \delta(\tau)$, $<n(\vec{r}_0,t)n(\vec{r}_0+\vec{r}_{d},t)>= \delta(\vec{r}_{d})$ (here, the brackets $<...>$ denote the mean value), which means that the correlation time of the source $n(x,y,t)$ equals zero and the noise signal values $n(x,y,t)$ at any different points ($x_1$, $y_1$) and ($x_2$, $y_2$) are statistically independent. 

Physically, random spatial component $\xi(x,y)$ can be included into the green illumination by adding an electronically-addressed spatial light modulator that spatially modifies the green illumination. In such coloured noise, the spatial random illumination is characterised by a finite temporal correlation determined by the parameter $\tau_{c}$. It is assumed in the following that the noise correlation time $\tau_{c}$ is much smaller than the OASLM's response time $\varepsilon$. In addition, all instantaneous values $\xi(x,y,t)<-I_{0\text{g}}$ are changed to $\xi(x,y,t)=-I_{0\text{g}}$ since the summary green light intensity $I_{0\text{g}}+\xi(x,y,t)$ cannot be negative. Finally, the stochastic spatial model of the setup in Fig. \ref{fig1} takes the form
\begin{equation}
\label{spatial_stochastic_model_phase_modulation_conv}
\begin{array}{l}
\varepsilon \dfrac{d\Gamma(x,y)}{dt}=-\Gamma(x,y)\\
+\dfrac{1}{\alpha_{\text{b}} I_{\text{b}}(x,y)+ \tilde{\alpha}_{\text{g}} (I_{0\text{g}}+\xi(x,y))+\beta}+\gamma,\\
\tau_{c}\dfrac{d\xi(x,y)}{dt}=-\xi(x,y)+\sqrt{2D_{\text{g}}\tau_{c}} n(x,y,t),
\end{array}
\end{equation}
where all the Jones vector component determining the blue light intensity are the same as in Eq. (\ref{spatial_model_phase_modulation_conv}).

%%%%%%%%%%%%%%%%%%%%%%% FIG 5 %%%%%%%%%%%%%%%%%%%%%%%%
\begin{figure*}[t!]
\centering
\includegraphics[width=0.75\textwidth]{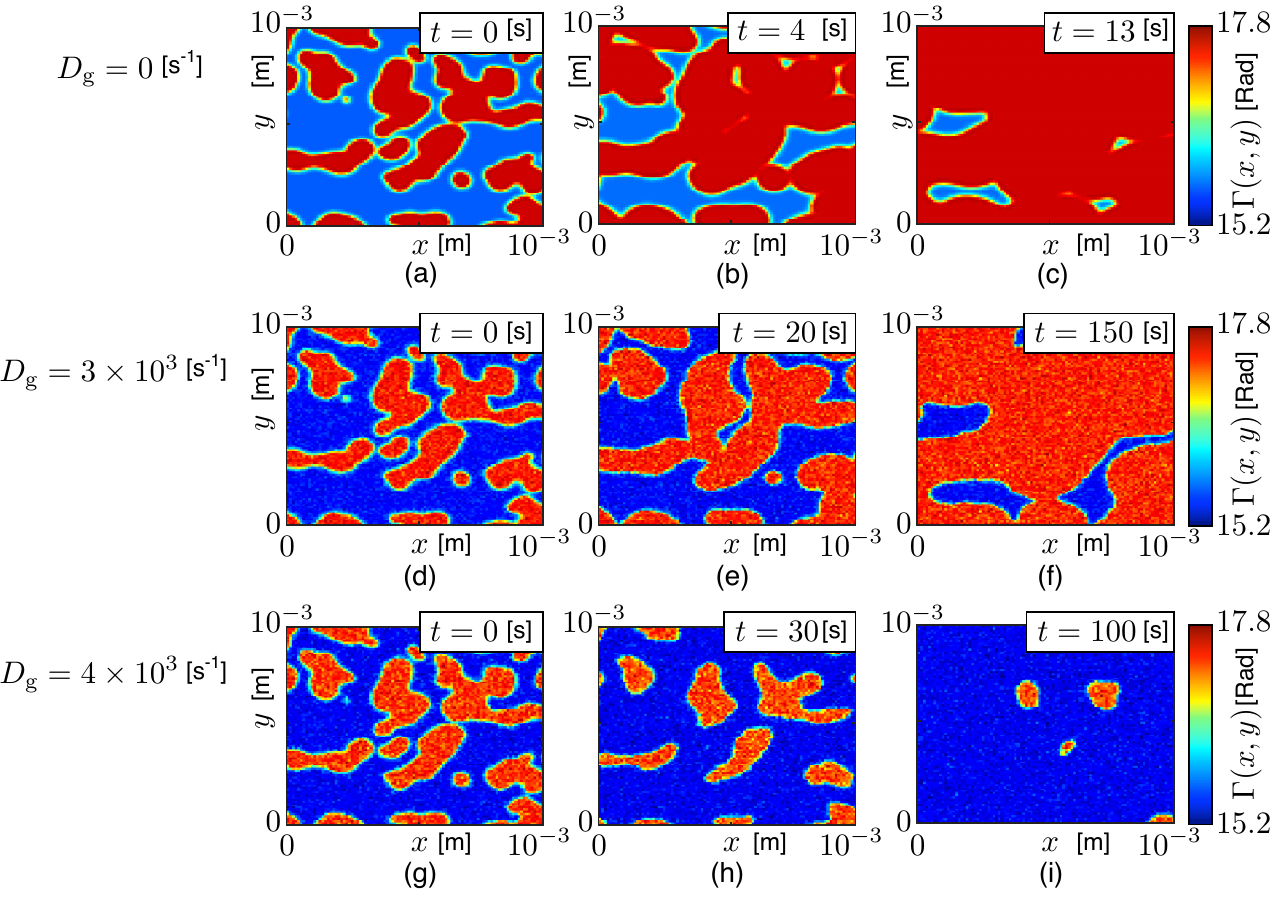}\\
\caption{Coarsening in Eq. (\ref{spatial_stochastic_model_phase_modulation_conv}) for increasing noise intensity: $D_{\text{g}}=0$ (panels (a-c)), $D_{\text{g}}=3\times 10^3$ (panels (d-f)),  $D_{\text{g}}=4\times 10^3$ (panels (g-i)). Other parameters are:  $\varepsilon=1$ [s], $\alpha_{\text{b}}=0.117$ [m$^2$Rad$^{-1}$W$^{-1}$], $\alpha_{\text{g}}=98.5\times 10^{-6}$ [m$^2$Rad$^{-1}$W$^{-1}$], $\beta=0.052$ [Rad$^{-1}$], $\gamma=-0.55$ [Rad], $\phi_0=\pi/2$ [Rad], $\phi_1=\pi$ [Rad], $\lambda_{\text{b}}=450\times10^{-9}$ [m], $\lambda_{\text{g}}=532\times10^{-9}$ [m], $R=0.95$, $I_{0\text{b}}=|\vec{E}_0|^2=0.01506$ [W/m$^2$], $I_{0\text{g}}=22$ [W/m$^2$], $\sigma=10^{-5}$ [m], $\tau_c=0.01$ [s].}
\label{fig5}
\end{figure*}  
%%%%%%%%%%%%%%%%%%%%%%%%%%%%%%%%%%%%%%%%%%%%%%%%%%%

First, Eq. (\ref{spatial_stochastic_model_phase_modulation_conv}) are considered for a set of parameters corresponding to Fig. \ref{fig3} (a) when the basin of attraction of steady state B is larger than the basin of state A. Equation (\ref{spatial_stochastic_model_phase_modulation_conv}) therefore exhibits coarsening and the system state $\Gamma (x,y)=B$ invades the entire space in the absence of noise, $D_{\text{g}}=0$ (see Fig. \ref{fig5} (a-c)). However, increasing noise intensity $D_{\text{g}}$ slows down the effect of coarsening, see Fig. \ref{fig5} (d-f), and above a threshold at around $D\approx 3.7\times 10^3$ [s$^{-1}$], noise inverts the the front propagation dynamics and state A dominates, see Fig. \ref{fig5} (g-i).

\begin{figure*}[t!]
\centering
\includegraphics[width=0.75\textwidth]{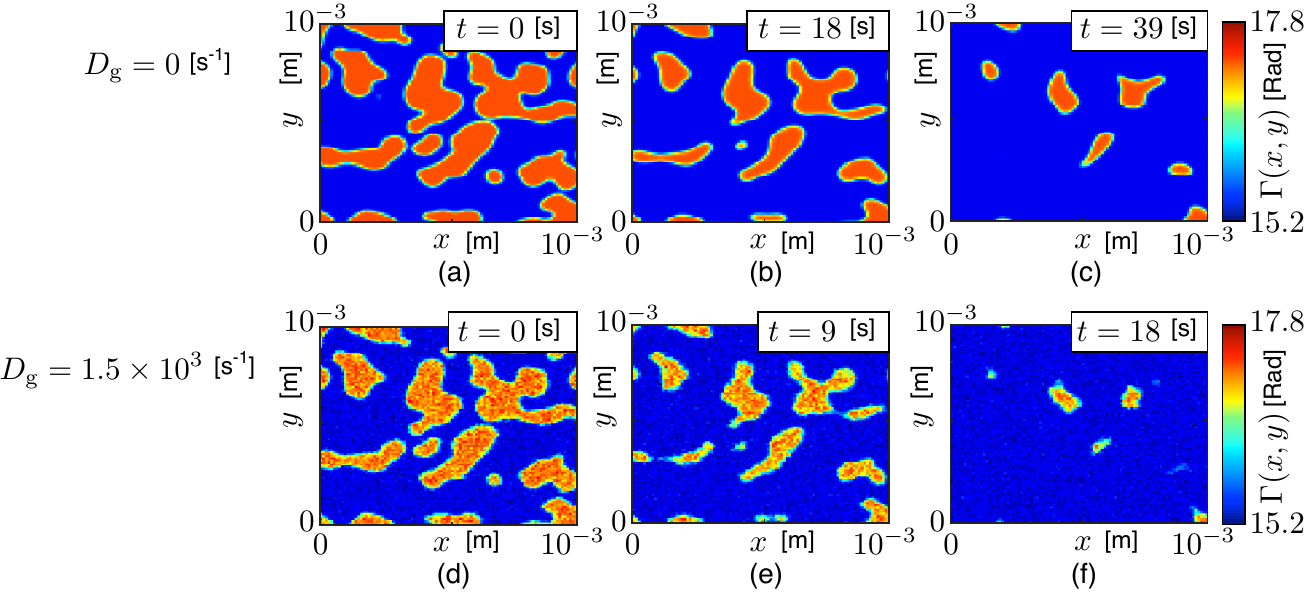}\\
\caption{Coarsening in Eq. (\ref{spatial_stochastic_model_phase_modulation_conv}) for increasing noise intensity: $D_{\text{g}}=0$ (panels (a-c)), $D_{\text{g}}=1.5\times 10^3$ (panels (d-f)). Other parameters are:  $\varepsilon=1$ [s], $\alpha_{\text{b}}=0.117$ [m$^2$Rad$^{-1}$W$^{-1}$], $\alpha_{\text{g}}=98.5\times 10^{-6}$ [m$^2$Rad$^{-1}$W$^{-1}$], $\beta=0.052$ [Rad$^{-1}$], $\gamma=-0.55$ [Rad], $\phi_0=\pi/2$ [Rad], $\phi_1=\pi$ [Rad], $\lambda_{\text{b}}=450\times10^{-9}$ [m], $\lambda_{\text{g}}=532\times 10^{-9}$ [m], $R=0.95$, $I_{0\text{b}}=|\vec{E}_0|^2=0.01506$ [W/m$^2$], $I_{0\text{g}}=36$ [W/m$^2$], $\sigma=10^{-5}$ [m], $\tau_c=0.01$ [s].}
\label{fig6}
\end{figure*}

Similarly, if the system parameter set corresponds to Fig. \ref{fig3} (e), one observes invading state A [Fig. \ref{fig6} (a-c)]. In such a case increasing the noise intensity speeds up the process [Fig. \ref{fig6} (d-f)]. Thus, it is demonstrated in Fig. \ref{fig5} and Fig. \ref{fig6} that, depending on the particular system configuration, noise can speed up coarsening, slow it down or even to invert the direction.

%%%%%%%%%%%%%%%%%%%%%%% FIG 6 %%%%%%%%%%%%%%%%%%%%%%%%
%%%%%%%%%%%%%%%%%%%%%%%%%%%%%%%%%%%%%%%%%%%%%%%%%%%

\begin{figure}[t]
\centering
\includegraphics[width=0.35\textwidth]{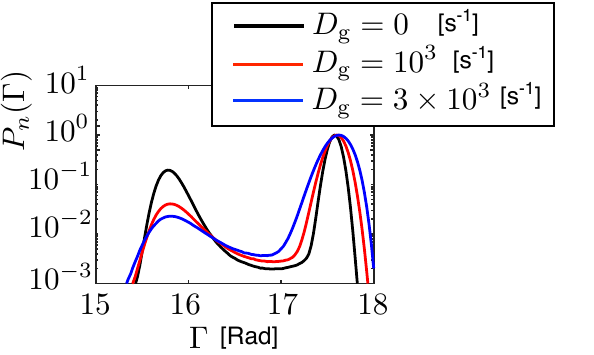}\\
\caption{Evolution of the normalised probability density function $P_{n}(\Gamma)$ caused by the varying noise intensity $D_{\text{g}}$ in Eq. (\ref{single_stochastic_PS_OASLM_phase_blue_green}). Parameters are:  $\varepsilon=1$ [s], $\alpha_{\text{b}}=0.117$  [m$^2$Rad$^{-1}$W$^{-1}$], $\alpha_{\text{g}}=98.5\times 10^{-6}$  [m$^2$Rad$^{-1}$W$^{-1}$], $\beta=0.052$ [Rad$^{-1}$], $\gamma=-0.55$ [Rad], $\phi_0=\pi/2$ [Rad], $\phi_1=\pi$ [Rad], $\lambda_{\text{b}}=450\times10^{-9}$ [m], $\lambda_{\text{g}}=532\times10^{-9}$ [m], $R=0.95$, $I_{0\text{b}}=0.01506$ [W/m$^2$], $I_{0\text{g}}=22$ [W/m$^2$], $\tau_c=0.01$ [s].}
\label{fig7}
\end{figure}

The theoretically rigorous explanation of the stochastic coarsening control in OASLM-based spatial models is significantly more challenging as compared with, for instance, the theoretical analysis given in Refs. \cite{garcia-ojalvo1999,mendez2011} for basic reaction-diffusion models with multiplicative noise. In particular, the 'small-noise-expansion approach' used in \cite{garcia-ojalvo1999,mendez2011} cannot be applied in the context of Eq. (\ref{spatial_stochastic_model_phase_modulation_conv}) due to the fact that any polynomial expression of Eq. (\ref{spatial_stochastic_model_phase_modulation_conv}) is challenging to obtain, and will furthermore give rise to stochastic terms in all the polynomial components. Consequently, it becomes extremely difficult to distinguish the systematic part of the noise influence. Nevertheless, we would like to emphasize the similarity between the processes observed in the basic models discussed in Refs. \cite{engel1985,garcia-ojalvo1999,mendez2011} and in OASLM-based spatial model described by Eq. (\ref{spatial_stochastic_model_phase_modulation_conv}). To visualise the fact that stochastic forcing has an asymmetric impact on Eq. (\ref{spatial_stochastic_model_phase_modulation_conv}), a single-oscillator stochastic model corresponding to Eq. (\ref{spatial_stochastic_model_phase_modulation_conv}) at $\sigma \to 0$ is taken into consideration. If $\sigma \to 0$, the spatial coupling is absent and the retardation $\Gamma$ individually evolves according to Eq. (\ref{single_oscillator}) at each point of the illuminated area, but in the presence of the noise term $\xi$ 
\begin{equation}
\label{single_stochastic_PS_OASLM_phase_blue_green}
\begin{array}{l}
\varepsilon \dfrac{d\Gamma}{dt}=-\Gamma+\dfrac{1}{\alpha_{\text{b}} I_{\text{b}}+\tilde{\alpha}_{\text{g}} (I_{0\text{g}}+\xi)+\beta}\\
+\gamma+\sqrt{0.02}n_a(t),\\
\tau_{c}\dfrac{d\xi}{dt}=-\xi+\sqrt{2D_{\text{g}}\tau_{c}} n(t),\\
I_{\text{b}}=I_{0\text{b}}\Big\{1+R^2+ 2R\cos(2\phi_0+\phi_1+2\Gamma)       \Big\},
\end{array}
\end{equation}
where the additive white Gaussian noise term $\sqrt{0.02}n_a(t)$ has no impact on the system's symmetry and is included to obtain a stationary distribution of the normalised probability density function for the dynamical variable, $P_{n}(\Gamma)$, in numerical simulations. The evolution of $P_{n}(\Gamma)$ caused by increasing noise intensity $D_{\text{g}}$ illustrated in Fig. \ref{fig7} indicates that the left peak becomes broadened faster than the right one. Thus, the action of noise $\xi(t)$ is significantly stronger in the vicinity of the left steady state $\Gamma_{*}=A$. This effect is similar to the noise-induced evolution of $P_n(u)$ in the phenomenological model defined by equation $\dfrac{dx}{dt}=-x(x-a+\xi_a)(x+b+\xi_b)+\nabla^2 x$ (see paper \cite{engel1985}).
  
%%%%%%%%%%%%%%%%%%%%%%% FIG 7 %%%%%%%%%%%%%%%%%%%%%%%%
%%%%%%%%%%%%%%%%%%%%%%%%%%%%%%%%%%%%%%%%%%%%%%%%%%%

\section{Conclusions} 

The peculiarities of the bifurcation transitions to the bistable dynamics discussed in paper \cite{semenov2021} in the context of single-oscillator models, are reflected in the behaviour of the corresponding spatially-extended systems, as for example in Eq. (\ref{spatial_model_phase_modulation_conv}) or similar models corresponding to different OASLM's rotation angles or incident light polarization states, as formation of localized spatial domains corresponding to the attraction of two coexisting steady states. If the right-hand side function is asymmetric, the steady state characterized by the larger basin of attraction, invades the entire space. This process is accompanied by the effect of coarsening, which is determined by both asymmetry and the shape of evolving domains. 

Applying the saddle-node or pitchfork bifurcation conditions, one can remove the system asymmetry and then the dominating domain expansion is slowed down. Moreover, if the incident green and blue light intensities vary and obey the saddle-node bifurcation condition, one can controllably invert the front propagation direction. However, the saddle-node bifurcation conditions do not allow to rigorously define the absolutely symmetric state, while applying the pitchfork bifurcation conditions provide for mathematical derivation of appropriate parameter values. 

The second approach to control coarsening is the introduction of noise into the system. In particular, the presence of parametric noise modulating the green light intensity gives rise to speeding up or slowing down and inverting the effects of front propagation and coarsening. The ability to control the dynamics by increasing noise intensity strength results from the the fact that fluctuation growth changes the system symmetry. Detailed theoretical analysis of the stochastic control represents an issue for further investigations.

The interdisciplinary significance of the obtained results consists in developed approach for the control of propagating fronts in bistable spatially-extended systems of any nature exhibiting the coexistence of two steady states. Representing the function being responsible for the local dynamics in a polynomial form by using the Taylor-series expansion, one can derive the pitchfork or saddle-node bifurcation conditions in the similar way as in the current paper and then apply them to tune the systems's symmetry and, resultantly, the front propagation speed and direction. Finally, the developed approach for controlling the symmetry of bistable spatially-extended systems offers great opportunities for future implementations of spin-state networks. 

\section*{Acknowledgments}
We are very grateful to professor Serhiy Yanchuk for fruitful discussions.
This work has been supported by the EIPHI Graduate School (Contract No. ANR-17-EURE-0002) and the Bourgogne-Franche-Comt\'e Region and H2020 Marie Sk\l{}odowska-Curie Actions (MULTIPLY, No. 713694).
Results of numerical simulations presented in Sec. III and IV are obtained by V.V.S. in the framework Russian Science Foundation Grant No. 22-72-00038. 

%V.V.S. carried out numerical simulations of Eqs. (\ref{spatial_model_phase_modulation_conv}) and (\ref{spatial_stochastic_model_phase_modulation_conv}) in the framework Russian Science Foundation Grant No. 22-72-00038. 

\section*{Declarations}
\begin{itemize}
\item The authors have no conflicts to disclose.
\item The data that support the findings of this study are available from the corresponding author
upon reasonable request.
\end{itemize}

%%\bibliographystyle{apsrev4-1}
%%\bibliography{bibliography}

%merlin.mbs apsrev4-1.bst 2010-07-25 4.21a (PWD, AO, DPC) hacked
%Control: key (0)
%Control: author (72) initials jnrlst
%Control: editor formatted (1) identically to author
%Control: production of article title (-1) disabled
%Control: page (0) single
%Control: year (1) truncated
%Control: production of eprint (0) enabled
%

\end{document}